\documentclass[preprint,12pt,authoryear]{elsarticle}

\usepackage{amssymb}
\usepackage{color}
\usepackage{bm}        
\usepackage{latexsym,amsmath}
\usepackage{dcolumn}   
\usepackage{natbib}
\journal{arXiv.org}

\begin{document}

\begin{frontmatter}



\title{Extreme control of impulse transmission by cylinder-based nonlinear phononic crystals}


\author[1]{Rajesh Chaunsali}
\author[2]{Matthew Toles}
\author[1]{Jinkyu Yang}
\author[1,3,4]{Eunho Kim \corref{5}}

\address[1]{Aeronautics and Astronautics, University of Washington, Seattle, WA, USA, 98195-2400}
\address[2]{Materials Science and Engineering, University of Washington, Seattle, WA, USA, 98195-2120}
\address[3]{Division of Mechanical System Engineering, 567 Baekje-daero, Deokjin-gu, Jeonju, Jeonbuk, Republic of Korea, 54896}
\address[4]{Automotive Hi-Technology Research Center \& LANL-CBNU Engineering Institute-Korea, Chonbuk National University, 567 Baekje-daero, Deokjin-gu, Jeonju, Jeonbuk, Republic of Korea, 54896}

\cortext[5]{Author to which all correspondence should be addressed: eunhokim@jbnu.ac.kr}

\begin{abstract}
We present a novel device that can offer two extremes of elastic wave propagation --- nearly complete transmission and strong attenuation under impulse excitation. The mechanism of this highly tunable device relies on intermixing effects of dispersion and nonlinearity. The device consists of identical cylinders arranged in a chain, which interact with each other as per nonlinear Hertz's contact law. For a `dimer' configuration, i.e., two different contact angles alternating in the chain, we analytically, numerically, and experimentally show that impulse excitation can either propagate as a localized wave, or it can travel as a highly dispersive wave. Remarkably, these extremes can be achieved in this periodic arrangement simply by \textit{in-situ} control of contact angles between cylinders. We close the discussion by highlighting the key characteristics of the mechanisms that facilitate strong attenuation of incident impulse. These include low frequency to high frequency (LF-HF) scattering, and turbulence-like cascading in a periodic system. We thus envision that these adaptive, cylinder-based nonlinear phononic crystals, in conjunction with conventional impact mitigation mechanisms, could be used to design highly tunable and efficient impact manipulation devices.   
\end{abstract}

\begin{keyword}
Phononic crystal \sep nonlinear \sep Hertz contact \sep stress wave \sep solitary wave \sep tunability \sep impact mitigation 
\end{keyword}

\end{frontmatter}

\section{Introduction}
\label{}
The ability to control the propagation of elastic waves through a structure is desired in numerous engineering applications, such as impact and shock absorption, wave filtering and isolation, energy harvesting, structural health monitoring, and nondestructive evaluations. Recent advances in the field of mechanical metamaterials and phononic crystals, which aim to achieve unconventional dynamic properties by artificially designing a mechanical structure, have opened up exciting possibilities to tailor elastic wave propagation as per given engineering requirements \citep{Kadic2013, Hussein2014}. 

A granular crystal, one such type of phononic crystal, is a systematic arrangement of granules in a closely packed form. These have become popular test-beds of various wave dynamics phenomena in recent years~\citep{Nesterenko2001, Sen2008, Porter2015}. What makes them special is their tunability, which stems from the nonlinear Hertz contact law \citep{Johnson1985} between granules. That is, for a given input excitation, the initial static compression in the system can be easily adjusted to invoke the wave dynamics to be nearly linear, weakly nonlinear, or strongly nonlinear. In this regard, sphere-based granular crystals have been commonly used for investigating various wave phenomena. There have also been studies on granular crystals consisting of both spheres and cylinders, which demonstrated additional tunability in terms of controlling particle mass by changing cylinder length without changing contact stiffness \citep{Nesterenko2001, Herbold2009}. More recently, phononic crystals based solely on cylinders have emerged as an alternate highly tunable architecture in terms of their wave propagation characteristics. These have several unique features, including -- but not limited to -- the control of contact stiffness based on the stacking angles of cylinders \citep{Khatri2012}, and the utilization of cylinder lengths to induce local resonance effects due to bending \citep{Kim2014, Kim2015a, Kim2015b}. Consequently, such systems can support a wide range of wave tailoring functionality, encompassing from filtering linear elastic waves by the formation of frequency band-gaps \citep{Kim2014, Li2012, Meidani2015, Chaunsali2016} to manipulating highly nonlinear waves \citep{Khatri2012, Kim2015a, Kim2015b, Kore2016}.

In this research, we explore the wave tailoring functionality of the cylinder-based nonlinear phononic crystal, relying solely on the first aforementioned feature, i.e., control over the contact stiffness by varying the stacking angles of cylinders under impulse excitation.  In particular, we explore initially uncompressed one-dimensional `dimer' with two contact angles periodically alternating along the chain. This thereby results in an effective system with periodic variation of `soft' and `stiff' contacts. We show that such a periodic system possesses two seemingly opposite extremes of wave phenomena: \textit{localized} traveling wave resulting in highly efficient impulse transmission, and highly \textit{dispersive} wave resulting in strong attenuation of the impulse. 

Though various studies on sphere-based, strongly nonlinear granular arrangements have shown promising use of intermixing nonlinearity with dispersion to achieve impulse transmission or attenuation \citep{Nesterenko2001, Hong2005, Daraio2006, Doney2006, Porter2008, Jayaprakash2011, Jayaprakash2013, Potekin2013, Leonard2014, Kim2015c}, there are still very limited architectures, which offer \textit{in-situ} control of highly nonlinear impulse excitation \citep{Pal2014}. We show analytically, numerically, and experimentally that the cylinder-based dimer can offer extreme control over the wave transmission characteristics simply by altering the stacking angles of the cylinders, which affects the interplay of nonlinearity and dispersion in the system. 

We also show that, unlike the energy absorbing mechanisms in conventional systems (e.g., foams, blankets, and disordered granular beds) generally showing large dissipation, the attenuation of impulse in the proposed system is achieved without resorting to any material dissipation. Rather, the attenuation mechanism is related to the energy redistribution process over the space- and time-domains of the system. Specifically, we find that this system leverages low- to high-frequency energy transfer and low- to high-wavenumber cascading mechanism across different time- and length-scales \citep{Kim2015c}. This is phenomenologically analogous to turbulence in fluids, and such mechanisms can make the cylinder-based nonlinear phononic crystals efficient in absorbing impact even in the absence of material dissipation.
  
The rest of the manuscript is arranged in the following manner. In Section 2, we write the governing equations of motion, and show that the qualitative nature of the wave dynamics in a dimer cylinder chain depends only on one system parameter: the ratio of stiffness coefficients of stiff and soft contacts. In Section 3, we perform extensive numerical simulations to show the existence of multiple stiffness coefficient ratios, which support localized traveling waves. In Section 4, we perform approximate asymptotic analysis to get insights into the system dynamics, and show that there would exist multiple such stiffness coefficient ratios, which invoke wave localization.  In Section 5, we include experimental demonstration of the existence of localized pulses, and the corresponding effect on the force transmission profiles. Additionally, we detect the contact angles, at which dispersion effects are maximum. These reflect as local minima in the force transmission curve indicating strong pulse attenuation. In Section 6, we present some critical insights into this dissipation-independent, passive mechanism that facilitates strong attenuation of input impulse. In Section 7, we conclude the study and suggest potential future applications.

\section{System description}
\begin{figure}[t]
\centering
\includegraphics[scale=0.5]{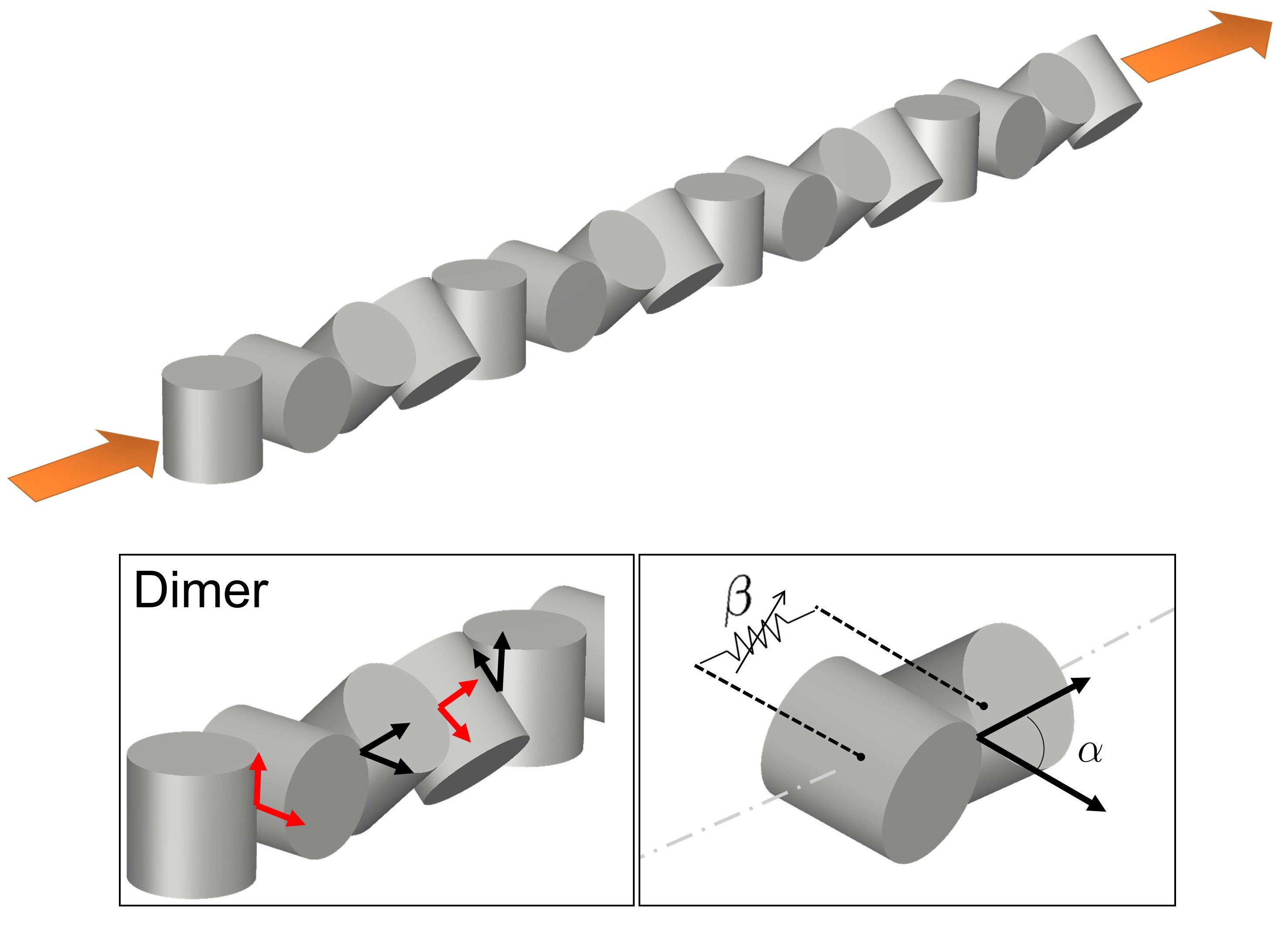}
\caption{\textbf{Cylinder-based nonlinear phononic crystal}. A one-dimensional chain, which allows manipulation of elastic waves by changing the contact angles of cylinders. (Left inset) A dimer chain with two alternating contact angles denoted by different colors. (Right inset) By changing the contact angle $\alpha$, one can change the stiffness coefficient $\beta$ between two cylinders. }
\label{fig1}
\end{figure}  

The system consists of identical short cylinders stacked one-dimensionally (Fig.~\ref{fig1}). These cylinders interact as per Hertz contact law and provide nonlinear stiffness variation along the chain. Each cylinder is free to be rotated so that one can appropriately set the angle between any two cylinders (further details of experimental setup to be described in Section 5). For this study, we construct a dimer system by choosing two different contact angles varying periodically along the chain. This variation is shown in the left inset of  Fig.~\ref{fig1} by two distinct colors. Owing to the geometry of cylinders, the contact angle $\alpha$ dictates the contact stiffness coefficient $\beta$ as shown in the right inset of Fig.~\ref{fig1}. Hence, by changing the contact angles, we can vary the contact stiffness coefficients, and thus influence the wave propagation along the chain.   

Mathematically, the contact force between $i$-th and $(i+1)$-th cylinders can be written as $\beta_i(\alpha) \delta^{3/2}$, with $\delta$ as the compression between two cylinders. Stiffness coefficient $\beta_{i}(\alpha)$, which is a function of the contact-angle $\alpha$,  is of the following form \citep{Johnson1985}:
\begin{eqnarray}
\begin{aligned}
\label{eqn1}
\beta_i(\alpha) =\frac{2Y}{3(1-\nu^2)} &\sqrt{\frac{R}{\sin{\alpha}}} \Bigg[\frac{2K_{e}(\alpha)}{\pi}\Bigg]^{-3/2} .\\
&\Bigg\{\frac{4}{\pi e({\alpha})^2} \sqrt{\bigg[\bigg(\frac{a}{b}\bigg)^2 E_{e}(\alpha)-K_{e}(\alpha)\bigg][K_{e}(\alpha)-E_{e}(\alpha)]} \Bigg\}^{1/2}. 
\end{aligned}
\end{eqnarray}

\noindent Here, $R$, $Y$, and $\nu$ denote the radius, Young's modulus, and Poisson's ratio of all identical cylinders. An elliptical contact area of two contacting cylinders has eccentricity $e(\alpha)=\sqrt{1-(b/a)^2}$, with $a$ and $b$ being the semi-major and semi-minor axes, respectively. The ratio $b/a$ can be approximated to $[(1+\cos{ \alpha})/(1-\cos{\alpha})]^{-2/3}$. $K_{e}(\alpha)$ and  $E_{e}(\alpha)$ are complete elliptical integrals of the first and second kinds, respectively. 

\begin{figure}[t]
\centering
\includegraphics[scale=0.5]{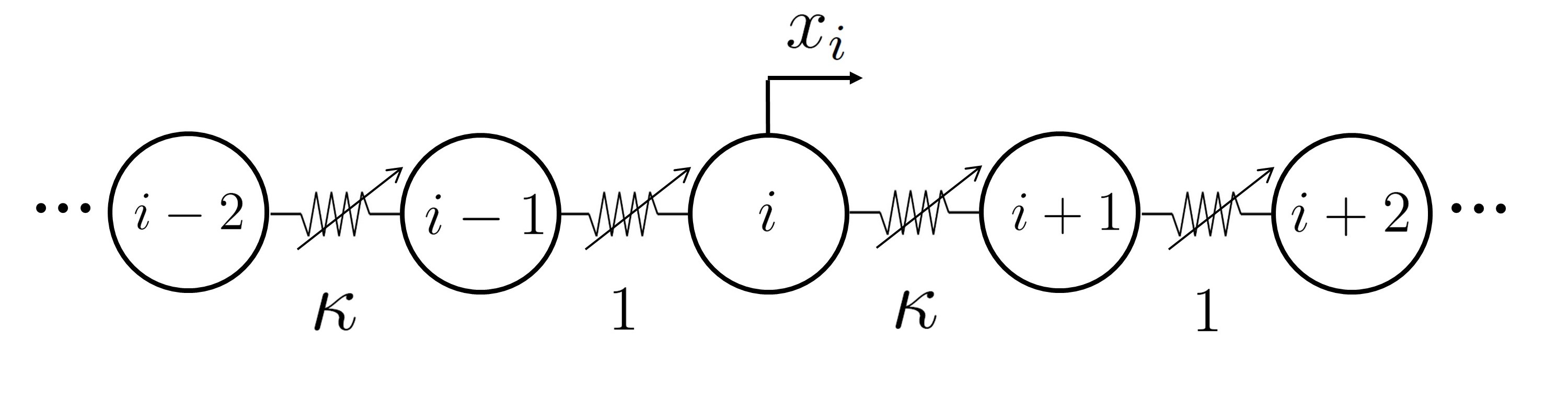}
\caption{Non-dimensional dimer chain with alternating soft (1) and stiff ($\kappa$) contact coefficients.}
\label{fig2}
\end{figure}  

We model the system using the discrete element method \citep{Cundall1979}, where each cylinder is represented by a lumped mass $m$ that is connected with neighboring particles by nonlinear springs following Hertz force law with the aforementioned coefficient $\beta_i(\alpha)$ (from this point on, we drop $\alpha$ for the brevity of expressions). Let $u_{i}$ denote the displacement of the $i$-th cylinder. In the absence of initial static compression to the chain, the dimer system can be represented by these equations of motion for two adjacent $i$-th and $(i+1)$-th cylinders- 
\begin{eqnarray}
\label{eqn2}
\begin{aligned}
m \frac{d^2 u_{i}}{dt^2} &= \beta_{i-1}(u_{i-1}-u_{i})^{3/2}_{+}
 - \beta_{i}(u_{i}-u_{i+1})^{3/2}_{+}, \\
m \frac{d^2 u_{i+1}}{dt^2} &= \beta_{i}(u_{i}-u_{i+1})^{3/2}_{+}
 - \beta_{i+1}(u_{i+1}-u_{i+2})^{3/2}_{+}. \\
\end{aligned}
\end{eqnarray}

\noindent Here, $(\Delta u)_{+}=max(\Delta u,0)$ is manifested due to the strictly compressive type of contact between cylinders. Note that any form of material dissipation is neglected in Eq.~\eqref{eqn2}, and $\beta_{i-1}=\beta_{i+1}$ by the definition of the dimer system. 

Using the radius of cylinders $R$ as the reference, we introduce the following non-dimensional parameters- 
\begin{eqnarray}
\label{eqn3}
x_{i}=\frac{u_{i}}{R}, ~~~ \tau=\Big [\frac{\beta_{i-1}R^{1/2}}{m} \Big]^{1/2} t
\end{eqnarray}

\noindent and deduce Eq.~\eqref{eqn2} into the non-dimensional equations of motion:
\begin{eqnarray}
\begin{aligned}
\label{eqn4}
\ddot{x}_i &= (x_{i-1}-x_{i})^{3/2}_{+} - \kappa(x_{i}-x_{i+1})^{3/2}_{+} \\ 
\ddot{x}_{i+1} &= \kappa(x_{i}-x_{i+1})^{3/2}_{+} -(x_{i+1}-x_{i+2})^{3/2}_{+}.
\end{aligned}
\end{eqnarray}
Here overdots denote a double derivative with respect to non-dimensional time $\tau$. $\kappa$ is the ratio of the stiffness coefficients in the dimer chain, which satisfies $ \kappa=\beta_{i}/\beta_{i-1}=\beta_{i}/\beta_{i+1} $, and so on. We note that $\kappa$, which depends on the alternating contact angles in the current system, is the characteristic parameter of the system. It dictates the qualitative nature of the dynamical response of the system. Thus we can simplify our system in the form of non-dimensional dimer, with alternating stiffness coefficients 1 and $\kappa$, and mass of each cylinder to be 1 (Fig.~\ref{fig2}). 

In order to analyze the entire family of dimer system, we would vary $\kappa$ in the interval $[1, \infty)$, such that, 1 and $\kappa$ denote the stiffness coefficients for `soft' and `stiff' contacts, respectively. The lower limit, i.e., $\kappa=1$ represents a `homogeneous' chain with all soft contacts with coefficient 1. In the upper limit, i.e., $\kappa \rightarrow \infty$, the two masses connected through $\kappa$, would tend to have perfectly rigid connection, and can be treated as one mass of two units. Therefore, the system again reduces to a homogeneous chain with these masses connected by only `soft' contacts (cf. Eq. (\ref{eqn12}) below). We will discuss more on this `auxiliary system' in later sections as the asymptotic method will tend to be accurate only close to this upper limit.    

\section{Numerical simulations}

\begin{figure}[t]
\centering
\includegraphics[scale=0.5]{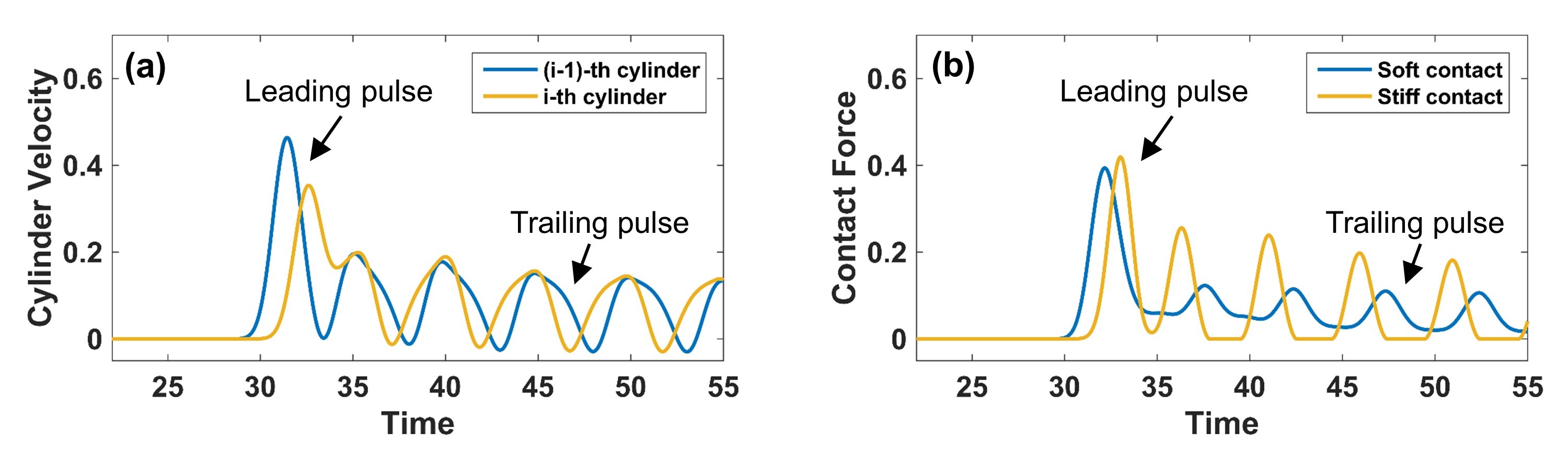}
\caption{\textbf{Typical propagating impulse with leading and trailing sections in non-dimensional dimer system.} (a) Time-history of the cylinder velocity at two adjacent cylinder locations. (b) Time-history of the contact force at two adjacent contacts.}
\label{fig3}
\end{figure}  
In order to numerically simulate the wave propagation through the dimer system, we employ the Runge-Kutta method and directly solve the non-dimensional Eq.~(\ref{eqn4}) for the entire system consisting of 300 cylinders. We fix the right end of the chain and apply a striker impact on the other end of the chain to generate a propagating impulse. The striker is identical to the cylinder used in the chain and has a unit striking velocity. We solve the system at multiple $\kappa$ values (i.e., $[1, \infty)$), and closely observe the profile of the propagating wave before it reaches the other end of the chain. 

Figure~\ref{fig3} depicts a typical time-history of the particles' velocity and the contact force at adjacent locations ($i=29,30$) in the chain. Given the two different stiffness coefficient values, in the ratio of $\kappa=2.398$ in this case, the system is no more a homogeneous system supporting localized waves \citep{Khatri2012}, and likely to have dispersive propagating waves. This is what we observe in the form of trailing pulse, which results from leading pulse shedding its energy. The dynamics of these trailing pulses is non-smooth as it involves loss of contacts (i.e., zero contact force in Fig.~\ref{fig3}b). This energy leakage thus hints an interesting interplay between nonlinearity and dispersion in the system.
Consequently, the system is passively able to attenuate an impulse as it travels along the chain. As we will see later, it is interesting that such an energy leakage could be \textit{maximized} at certain $\kappa$ values, and those accompany interesting energy redistribution mechanisms.

\begin{figure}[t]
\centering
\includegraphics[scale=0.6]{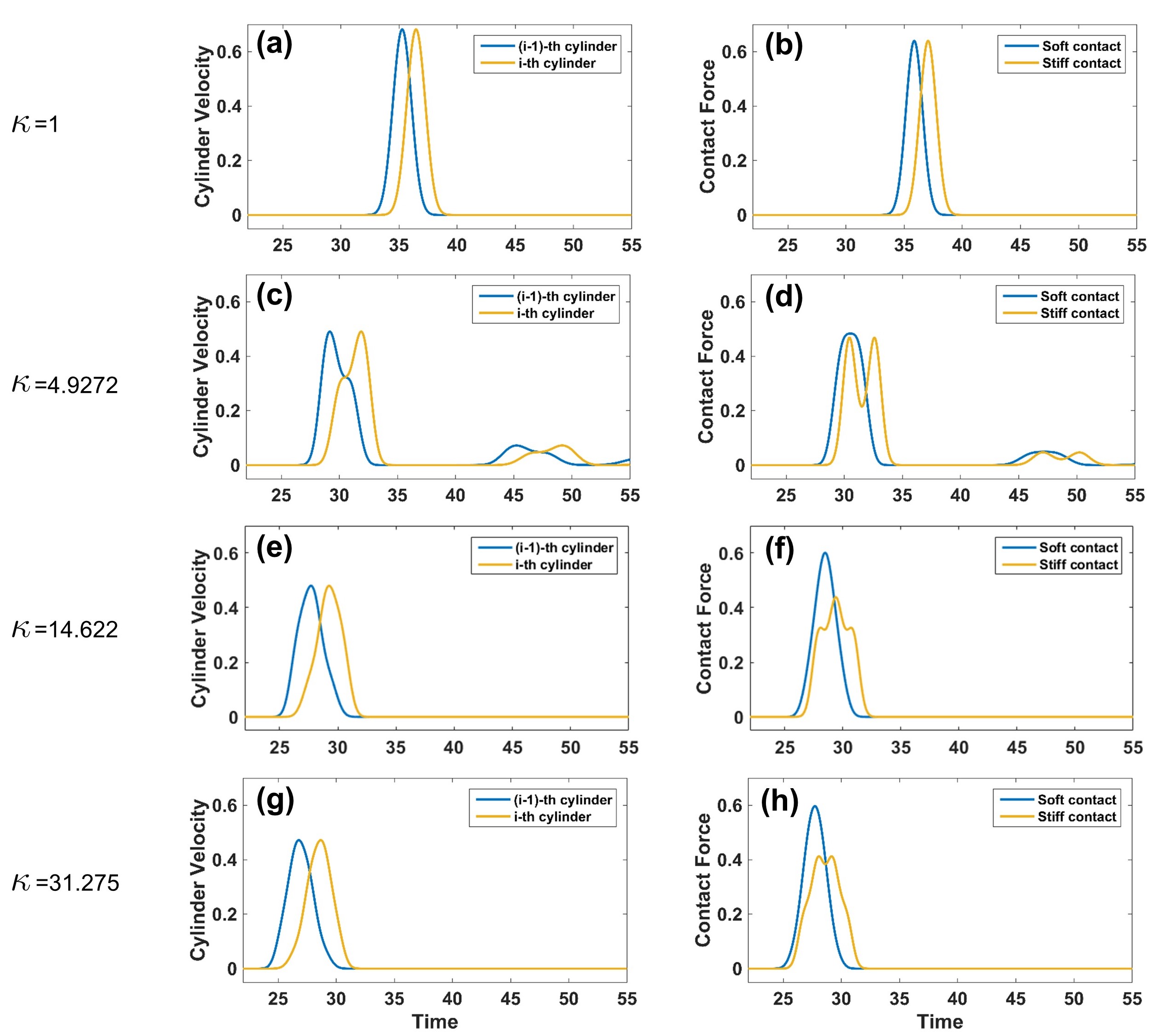}
\caption{\textbf{Localized propagating impulse waves at different stiffness coefficient ratios, $\kappa$.} (a,c,e,g) Time-history of the cylinder velocity at two adjacent cylinder locations. (b,d,f,h) Time-history of the contact force at two adjacent contacts.}
\label{fig4}
\end{figure}  

For some $\kappa$ values, however, the dimer system exhibits entirely different wave propagation characteristics. In such scenarios, as opposed to the expected trailing pulses emanating from the leading pulse, dispersion effect is balanced by nonlinearity, thereby resulting in \textit{complete} absence of these trailing pulses. This means that the energy in the form of propagating pulse does not leak into the wave tails as it travels along the system. Rather, the pulse is highly localized and moves forward with a uniform speed. 

In Fig.~\ref{fig4} we plot the time-history of particles' velocities and contact forces at the same locations as that in Fig.~\ref{fig3}. Upon increasing $\kappa$ from 1 to higher values, we have taken first four scenarios showing wave localization. We first note that there exist multiple $\kappa$, at which the dimer system supports localized traveling waves, leading to impulse transmission without any attenuation. Moreover, one can also notice that the time-history plots at adjacent locations are different. In particular, we note that contact force profiles at stiff contact shows higher fluctuations (more humps) compared to that at soft contact.  As one deviates from the homogeneous case ($\kappa=1$) by increasing the stiffness coefficient ratio, these fluctuations are enhanced but carry lesser amplitude.  

By further increasing the ratio, the fluctuations are expected to be infinitely higher but reduce to zero in amplitude. This makes sense because for infinitely large values of $\kappa$, the stiff contact can be approximated as perfectly rigid contact. This would make the effective system homogeneous with masses two times that of a cylinder, and those would interact through soft contacts. Consequently, the time-history plots for adjacent locations would qualitatively look alike as that in Fig.~\ref{fig4} a, b (for example, compare its profiles with high $\kappa$ case as in Fig.~\ref{fig4} g, h). The known solution of this auxiliary system would be used later for asymptotically solving the system near large values of $\kappa$. 

We would also like to point that a localized pulse of smaller amplitude traveling behind the main pulse can also be observed (see for example Fig.~\ref{fig4} c, d). This is because of the separation and multiple collisions among the initial masses. We observe such waves, of varying amplitudes though, in all the cases except $\kappa$=1.  This does not contradict the aforementioned localization behavior in the system, as initially generated localized pulses \textit{still} travel in the medium without any attenuation. Rather, it further corroborates that the stiffness coefficient ratio fundamentally characterizes the system in the sense that impulse of any amplitude (within elastic limits) would travel along the system in a localized fashion. 

These localized pulses are closely related to --- so called `solitary waves' --- found in granular dimer systems with alternating masses \citep{Jayaprakash2011}. However, in the strict sense, the localized waves found in our cylindrical system are different because of the \textit{asymmetry} in the velocity profiles (see Figs.~\ref{fig4} a, c, e, and g). Instead, the contact force profile (Figs.~\ref{fig4} b, d, f, and h) is symmetric.  This distinction is important to highlight because of two reasons. First, it contributes to our general understanding of the various families of solitary waves, which an inhomogeneous strongly nonlinear medium (dimer) can support. The presented solitary wave deviates from the conventional notion that the family of solitary wave would have a symmetric velocity profile as observed in dimer chains having different particle masses \citep{Jayaprakash2011}. Second, it is the special symmetry condition satisfied by the contact force, which will eventually be used for the asymptotic method later to analytically estimate $\kappa$ values supporting the solitary wave.

\section{Analytical study}

We now employ an analytical method to predict the aforementioned stiffness coefficient ratios ($\kappa$), which ensure the localization of propagating waves in the system in the form of traveling solitary waves. To this end, we use the asymptotic method, whose essence lies in separating different time-scales in system dynamics \citep{Bender1999}. We would show below that asymptotic analysis would be accurate when $\kappa \rightarrow \infty$. This case, as already mentioned before, is the auxiliary system, and represents a homogeneous chain supporting a solitary wave \citep{Nesterenko2001, Sen2008, Khatri2012}. We use the relation given in the reference \cite{Sen2008} to represent this solitary wave, and then approximate the dynamics for  $1 \ll\kappa<\infty$ by adopting the similar strategy employed by \cite{Jayaprakash2011} in the context of the sphere-based dimer chain.   

Rewriting the equations of motion from Eq.~(\ref{eqn4}) for three adjacent cylinders at $(i-1)$-th, $i$-th, and $(i+1)$-th locations
\begin{eqnarray}
\begin{aligned}
\label{eqn5}
\ddot{x}_{i-1} &= \kappa(x_{i-2}-x_{i-1})^{3/2}_{+} -(x_{i-1}-x_{i})^{3/2}_{+}, \\
\ddot{x}_i &= (x_{i-1}-x_{i})^{3/2}_{+} - \kappa(x_{i}-x_{i+1})^{3/2}_{+}, \\ 
\ddot{x}_{i+1} &= \kappa(x_{i}-x_{i+1})^{3/2}_{+} -(x_{i+1}-x_{i+2})^{3/2}_{+}. 
\end{aligned}
\end{eqnarray}

\noindent which can be rearranged as following after subtraction
\begin{eqnarray}
\begin{aligned}
\label{eqn6}
\ddot{x}_{i-1}-\ddot{x}_i &= \kappa(x_{i-2}-x_{i-1})^{3/2}_{+} -2(x_{i-1}-x_{i})^{3/2}_{+} + \kappa(x_{i}-x_{i+1})^{3/2}_{+},  \\
\ddot{x}_i-\ddot{x}_{i+1} &= (x_{i-1}-x_{i})^{3/2}_{+} - 2\kappa(x_{i}-x_{i+1})^{3/2}_{+} +(x_{i+1}-x_{i+2})^{3/2}_{+}.  \\ 
\end{aligned}
\end{eqnarray}

Let us define a quantity $\psi_i=F^{2/3}_i$, where $F_i$ denotes the contact force between $i$-th and $(i+1)$-th cylinders. Hence, for the current dimer system (Fig.~\ref{fig2}), $\psi_{i}=\kappa^{2/3}(x_{i}-x_{i+1})$, $\psi_{i+1}=(x_{i+1}-x_{i+2})$, and so on. Substituting these relations into Eq.~(\ref{eqn6}), we transform the equations of motion in terms of $\psi$ as
\begin{eqnarray}
\begin{aligned}
\label{eqn7}
\ddot{\psi}_{i-1}&= (\psi_{i-2})^{3/2}_{+} + (\psi_{i})^{3/2}_{+} -2(\psi_{i-1})^{3/2}_{+}, \\
\epsilon ~\ddot{\psi}_{i}&= (\psi_{i-1})^{3/2}_{+} + (\psi_{i+1})^{3/2}_{+} -2(\psi_{i})^{3/2}_{+} \\
\end{aligned}
\end{eqnarray}

\noindent where, $\epsilon=1/\kappa^{2/3}$. It is clear that $\epsilon \in (0,1]$ since $\kappa \in [1,\infty)$. Its multiplication with the overdot term in the second equation indicates that $\epsilon$ dictates a time-scale in the system. Therefore, we will see that for $\epsilon \ll 1$ (or $\kappa \gg 1$), time scales can be separated in the system. We seek asymptotic solution of the equations around this limit. Moreover, from this point on, we drop the subscript (+) as we focus only on the localized wave where cylinders do not lose contact.  

Expanding $\psi$ asymptotically (for small $\epsilon$) in the two distinct time-scales, $t_0$(slow) and $t_1$(fast), results in
\begin{eqnarray}
\begin{aligned}
\label{eqn8}
\psi_{i-1}&=\psi_{i-1,0}(t_0)+\epsilon^{\gamma}~\psi_{i-1,1}(t_1)+...~~(\text{soft contact})\\
\psi_{i}&=\psi_{i,0}(t_0)+\epsilon^{\alpha}~\psi_{i,1}(t_1)+...~~(\text{stiff contact}) \\
\text{for} \quad i &=... (i-2), i, (i+2)...   
\end{aligned}
\end{eqnarray}

\noindent where, $t_0=\tau$, $t_1=\epsilon^{\beta}~\tau$; and $\alpha$,$~\beta$, and $\gamma$ are the constants to be determined by balancing the terms in asymptotic expressions. For brevity, we drop to time-scale notation in parentheses, e.g., $\psi_{i,0}(t_0)=\psi_{i,0}$.

We substitute asymptotic expansions of Eq.~(\ref{eqn8}) in Eq.~(\ref{eqn7}), and collect the powers of $\epsilon$ to get 
\begin{eqnarray}
\begin{aligned}
\label{eqn9}
\ddot{\psi}_{i-1,0} + \epsilon^{\gamma+2\beta} \psi''_{i-1,1} &= \psi^{3/2}_{i-2,0} + (3/2)\psi^{1/2}_{i-2,0}~ \epsilon^{\alpha} \psi_{i-2,1}\\
&+ \psi^{3/2}_{i,0} + (3/2)\psi^{1/2}_{i,0}~ \epsilon^{\alpha} \psi_{i,1}\\
&- 2\psi^{3/2}_{i-1,0} - 3\psi^{1/2}_{i-1,0}~ \epsilon^{\gamma} \psi_{i-1,1} \\
&+ O(\Vert \epsilon^{\gamma} \psi_{i-1,1}\Vert^{2})+O(\Vert \epsilon^{\alpha} \psi_{i,1}\Vert^{2})
\end{aligned}
\end{eqnarray}

\begin{eqnarray}
\begin{aligned}
\label{eqn10}
\epsilon(\ddot{\psi}_{i,0} + \epsilon^{\alpha+2\beta} \psi''_{i,1}) &= \psi^{3/2}_{i-1,0} + (3/2)\psi^{1/2}_{i-1,1}~ \epsilon^{\gamma} \psi_{i-1,1}\\
&+ \psi^{3/2}_{i+1,0} + (3/2)\psi^{1/2}_{i+1,0}~ \epsilon^{\gamma} \psi_{i+1,1}\\
&- 2\psi^{3/2}_{i,0} - 3\psi^{1/2}_{i,0}~ \epsilon^{\alpha} \psi_{i,1} \\
&+ O(\Vert \epsilon^{\gamma} \psi_{i-1,1}\Vert^{2})+O(\Vert \epsilon^{\alpha} \psi_{i,1}\Vert^{2})
\end{aligned}
\end{eqnarray}

\noindent where, $\ddot{\psi}$ and $\psi''$ represent double derivatives of $\psi$ with respect to time $t_0$ and $t_1$, respectively. We choose $\alpha=1$, $\beta=-1/2$ and $\gamma=2$ to balance the powers of $\epsilon$ on both sides, and then proceed towards analyzing system dynamics for slow and fast time-scales.

\textbf{Slow dynamics:} In order to predict the slow dynamics of the system, we take the zeroth-order approximation of Eq.~(\ref{eqn9}) and Eq.~(\ref{eqn10})  by setting $\epsilon=0$, and obtain 
\begin{eqnarray}
\begin{aligned}
\label{eqn11}
\ddot{\psi}_{i-1,0} &= \psi^{3/2}_{i-2,0}+\psi^{3/2}_{i,0}-2\psi^{3/2}_{i-1,0}, \\
\psi^{3/2}_{i,0} &= \frac{\psi^{3/2}_{i-1,0}+\psi^{3/2}_{i+1,0}}{2}
\end{aligned}
\end{eqnarray}

The above equations are in terms of only the slow time-scale $t_0$. If we substitute the later equation into the first, we obtain the following after some algebraic rearrangements
\begin{eqnarray}
\begin{aligned}
\label{eqn12}
\ddot{\psi}_{i-1,0} &= \bigg [\frac{\psi^{3/2}_{i-3,0} - \psi^{3/2}_{i-1,0}}{2}\bigg ] - \bigg [\frac{\psi^{3/2}_{i-1,0} - \psi^{3/2}_{i+1,0}}{2}\bigg ], \\
\psi^{3/2}_{i,0} &= \frac{\psi^{3/2}_{i-1,0}+\psi^{3/2}_{i+1,0}}{2}
\end{aligned}
\end{eqnarray}

\noindent Here the first equation represents the slow dynamics entirely expressed in the form of $\psi$ defined at the soft contacts ($i-1, i+1, i+3...$). This makes sense because slow dynamics ($\epsilon=0$) represents the auxiliary system, in which, dynamics is governed by only soft contacts. Solitary wave solution of this homogeneous system with the lumped masses of 2 units, can be obtained from the reference \cite{Sen2008}. Therefore, we use cylinder displacement $S$ to calculate contact force $F$, and finally $\psi=F^{2/3}$, to write-
\begin{eqnarray}
\begin{aligned}
\label{eqn13}
\psi_{i-1,0}(t_{0})=S_{i-1}(\eta t_{0})-S_{i}(\eta t_{0}), ~ \eta=(1/2)^{1/2}
\end{aligned}
\end{eqnarray}

\noindent where,
\begin{eqnarray}
\begin{aligned}
\label{eqn14}
S(\xi)=A+&(A/2)\bigg[\text{tanh}\big[ (C_1(\xi/T)+ C_3(\xi/T)^3 +C_5(\xi/T)^5)/2 \big] -1 \bigg], \\
& C_1=2.39536, C_3=0.268529, C_5=0.0061347
\end{aligned}
\end{eqnarray}

\begin{figure}
\centering
\includegraphics[scale=0.059]{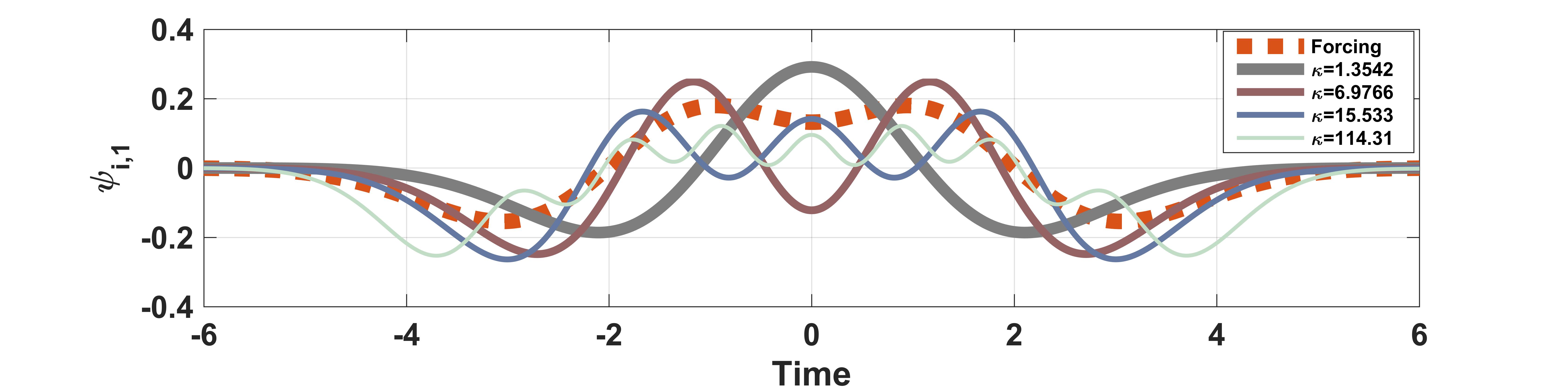}
\caption{Fast component solution of stiff contact dynamics $\psi_i$ as per Eq.~(\ref{eqn17}).  Multiple solutions are plotted at different stiffness coefficient ratio $\kappa (={1/\epsilon}^{3/2})$.} 
\label{fig5}
\end{figure} 

In Eq.~(\ref{eqn14}), $A$ denotes the amplitude of the solitary wave, and $T$ is the time-shift between the velocity peaks of any two alternate cylinders. Once we have explicit form of slow dynamics at soft contacts, we can easily obtain similar form for the stiff contacts from Eq.~(\ref{eqn12}) and Eq.~(\ref{eqn13}), and write 
\begin{eqnarray}
\begin{aligned}
\label{eqn15}
\psi_{i,0}(t_0) &= \frac{\big[S_{i-1,0}(\eta t_0) - S_{i,0}(\eta t_0)\big]^{3/2}}{2} + \frac{\big[S_{i,0}(\eta t_0) - S_{i+1,0}(\eta t_0)\big]^{3/2}}{2} \\
\end{aligned}
\end{eqnarray}

\textbf{Fast dynamics:} We now look into the fast dynamics of the system. This will essentially be linked to the higher oscillations observed at stiff contact in Fig.~\ref{fig4}. In order to calculate the analytical approximation, we take the first-order terms of $\epsilon$ from Eq.~(\ref{eqn9}) and Eq.~(\ref{eqn10}), and write
\begin{eqnarray}
\begin{aligned}
\label{eqn16}
\psi''_{i-1,1}(t_1)&=(3/2)\big[\psi^{1/2}_{i-2,0}(t_0)~ \psi_{i-2,1}(t_1) +\psi^{1/2}_{i,0}(t_0)~\psi_{i,1}(t_1)\big],\\
&\psi''_{i,1}(t_1)+ \Omega^{2}_i(t_0) \psi_{i,1}(t_1)=f_{i}(t_0)
\end{aligned}
\end{eqnarray}

\begin{table}
\begin{center}
\label{tab:param}
\begin{tabular}{cccc}
Order         & Direct numerical                               &  Numerical solution of                                       & Percentage    \\
 $n$              & simulation of Eq.~(\ref{eqn4})       &   asymptotically valid Eq.~(\ref{eqn17})             & Error            \\
\hline
\\
1 & 1.0000 & 1.3542 & 35.42\\
2 & 4.9272 & 6.9766 & 41.59\\
3 & 14.622 & 15.533 & 6.23 \\
4 & 31.275 & 29.809 & -4.69\\
5 & 54.517 & 50.130 & -8.05\\
6 & 82.990 & 77.948 & -6.08\\
7 & 119.92 & 114.31 & -4.68\\
18 & 1631.0 & 1629.9 & -0.07 \\
\end{tabular}
\caption{Comparison of stiffness coefficient ratio ($\kappa$) obtained through direct numerical simulation and asymptotic method for the formation of solitary waves.}
\end{center}	
\end{table} 

\noindent where, $\Omega^{2}_i(t_0)=3\psi^{1/2}_{i,0}(t_0)$, and $f_{i}(t_0)=-\ddot{\psi}_{i,0}(t_0)$. The second part of Eq.~(\ref{eqn16}) represents a linear oscillator; and clearly marks that the \textit{fast dynamics} of  the stiff contact parameter $\psi_{i}$ can be calculated if one knows its \textit{slow dynamics}. This is because frequency $\Omega_i$ and forcing $f_i$ of the oscillator solely depend on slow time-scale $t_0$. Fast dynamics of the soft contact can then easily be deduced from the first part of Eq.~(\ref{eqn16}) once we know fast dynamics of stiff contact. Therefore, for now, we would only focus on calculating the fast dynamics of the stiff contact.  Using the expressions for the two time-scales, $t_0=\tau$ and $t_1=\epsilon^{-1/2}\tau$, we can write the linear oscillator with a common, slow time-scale as
\begin{eqnarray}
\begin{aligned}
\label{eqn17}
\ddot{\psi}_{i,1}(\tau)+ \frac{\Omega^{2}_i(\tau)}{\epsilon} \psi_{i,1}(\tau)=\frac {f_{i}(\tau)}{\epsilon}
\end{aligned}
\end{eqnarray}

This equation is a second order differential equation, and we would now discuss the special symmetry conditions, which $\psi_{i,1}$ is required to follow to have solitary waves in the system. We noted earlier that $\psi$ has a specially symmetry because the contact force ($F=\psi^{3/2}$) was found to be symmetric relative to the point in time at which it attains maximum. By letting $\tau=0$ to be that reference point, we impose symmetry conditions directly on the fast component $\psi_{i,1}$ of $\psi_{i}$ as its slow component $\psi_{i,0}$ is already symmetric (Eq.~(\ref{eqn15})).  Therefore, to ensure wave localization with no trace of trailing pulses, we numerically solve the equation with the initial conditions $\psi_{i,1}(\tau \rightarrow -\infty)=0$ and $\dot{\psi}_{i,1}(\tau \rightarrow -\infty)=0$, and select the discrete cases resulting in symmetric profile of $\psi_{i,1}$, i.e., $\dot{\psi}_{i,1}(\tau \rightarrow \infty)=0$. 

These cases have associated eigenvalues $\epsilon_n$. In Fig.~\ref{fig5}, we summarize a few fast component solutions obtained at different  $\kappa~(={1/\epsilon_n}^{3/2})$ values. We first note that there are multiple such solutions, which agrees well with our earlier observations in numerical simulations. Moreover, as $\kappa$ increases, oscillations in the solution also increase. Once these solutions are superimposed on the slow solution (Eq.~(\ref{eqn15})) to get the overall dynamics at the stiff contact, these oscillations directly explain the `humps' in the time-history plot observed earlier (Fig.~\ref{fig4}). Therefore, to extend the argument, we can conjuncture that there would exist a countable infinity of solitary wave solutions obtained from Eq.~(\ref{eqn17}) with increasing degree of oscillatory nature. 

We use the number of maxima in $\psi_{i,1}$ profile as an index ($n$) to represent the order of solitary wave solution. Table 1 summarizes the comparison between the stiffness coefficient ratios obtained from numerical and asymptotic approaches for these indices. We see the percentage error oscillates initially, and start approaching 0 for $\kappa \gg 1$.

\section{Experimental verification}
\begin{figure}[t]
\centering
\includegraphics[scale=0.55]{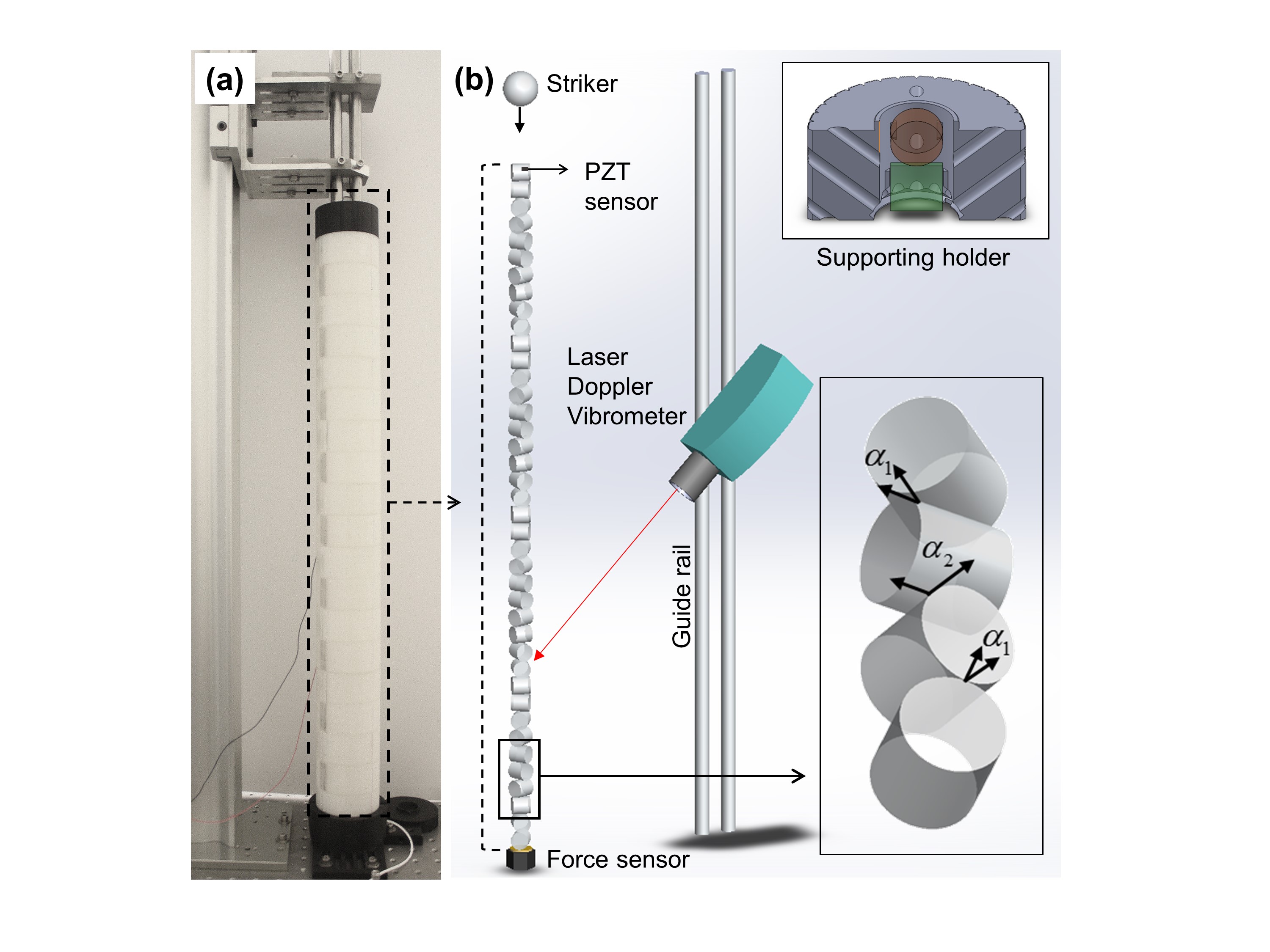}
\caption{\textbf{Experimental setup.} (a) The cylinder chain is aligned inside the cylindrical supporting holders. (b) Schematic diagram of the test setup. The upper inset shows a half-cut of cylindrical supporting holder including two orthogonal cylinders. The holes for the passage of laser beam can also be seen. The lower inset represents dimer chain with two alternating contact angles ($\alpha_1$, $\alpha_2$). }
\label{fig6}
\end{figure} 

Now we conduct experiments on the cylinder-based dimer system to validate the numerical and analytical results described in the previous sections. Figure~\ref{fig6} shows the experimental setup consisting of a cylinder chain, its supporting structure, and measurement systems. The chain is composed of 40 identical fused quartz cylinders (Young's modulus $Y = 72$ GPa, Poisson's ratio $\mu$ = 0.17, density $\rho = 2200$ kg/m$^3$). Each cylinder has its length and diameter equal to 18 mm. 

To align and hold the cylindrical chain, we fabricate 20 supporting holders using a 3D printer (upper right inset in Fig.~\ref{fig6}) and assemble them vertically. In each supporting holder, two quartz cylinders are positioned, and they maintain a contact angle of $\alpha_2 = 90^{\circ}$. We rotate the holders individually to maintain a constant angle ($\alpha_1$) between each one of them. As a results, we have 40 cylinder dimer system with periodically varying two different angles ($\alpha_1$, and $\alpha_2=90^{\circ}$, bottom inset in Fig.~\ref{fig6}). It is noted that there is some deliberate clearance made available for cylinders inside the holders to minimize the effect of friction on the propagating waves.

We drop a quartz sphere with a 18 mm diameter on the top of the chain with a 4 cm height to give an impulse excitation ($v_{impact}=0.88 m/s$). We measure the transmission with the force sensor (PCB 208 C02) located at the bottom of the chain. We also measure the dynamics of the system by using a non-contact laser Doppler vibrometer (LDV). We record particles' velocities by directing the laser beam to each particle through specially designed holes on the supporting holders (see the upper right inset in Fig.~\ref{fig6}b). We bond a small piezoelectric ceramic plate ($3 \times 4 \times 0.5$ mm) on the surface of the first cylinder. It generates a voltage signal at the moment of impact excitation, which is used for triggering the force sensor and the LDV system to record the measurement data. By using the combination of the contact force sensor and the non-contact LDV, we record the wave propagating properties of the system in various configurations. Specifically, we differentiate the angle sets ($\alpha_1$, $\alpha_2$) by varying $\alpha_1$ from $2^{\circ}$ to $90^{\circ}$ while keeping $\alpha_2$ fixed at $90^{\circ}$.

\begin{figure}[t]
\centering
\includegraphics[scale=0.55]{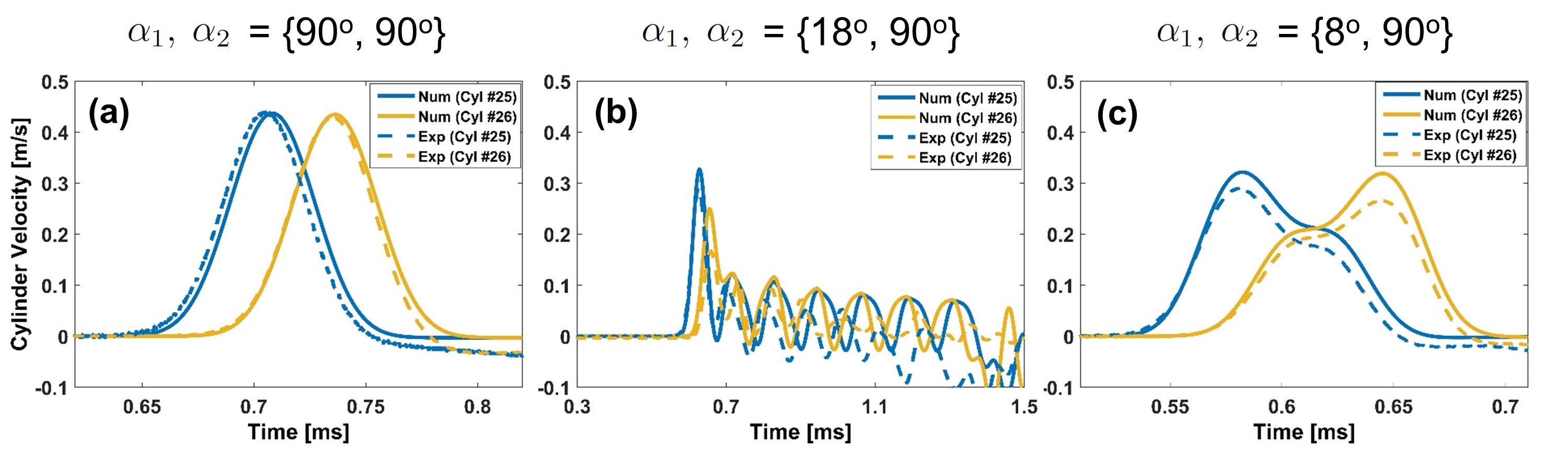}
\caption{\textbf{Comparing numerical and experimental results for the time-history of velocity for two adjacent cylinders.} (a) $\alpha_1=90^{\circ}$ (localized wave). (b) $\alpha_1=18^{\circ}$ (dispersive wave). (c) $\alpha_1=8^{\circ}$ (localized wave).} 
\label{fig7}
\end{figure} 

Figure~\ref{fig7} depicts the numerically and experimentally obtained velocities of particle number 25 and 26 in the chain for three different dimer configurations. The case when $\alpha_1, \alpha_2=\{90^{\circ}, 90^{\circ}\}$,  the chain represents a homogeneous chain, and therefore a localized solitary wave propagates in the medium as shown in Fig.~\ref{fig7}a. When we decrease $\alpha_1$ to 18$^{\circ}$, we see that the impulse wave sheds its energy in the form of oscillatory trailing waves as it propagates in the medium (Fig.~\ref{fig7}b). This is a case when dispersion effects dominate.  However, by further reducing $\alpha_1$ to 8$^{\circ}$, in Fig.~\ref{fig7}c we observe that the impulse wave localizes again, and resembles the asymmetric velocity profile in Fig.~\ref{fig4}c. The stiffness coefficient ratio $\kappa$ for this case can be calculated from Eq.~(\ref{eqn1}), and that equals 4.507. This is in the vicinity of numerically predicted value ($\kappa=4.9272$) for second order solitary wave formation (Table~1). Therefore, we confirm a trend of repeated localization of impulse as we decrease $\alpha_1$ (or increase $\kappa$).

\begin{figure}[t]
\centering
\includegraphics[scale=0.55]{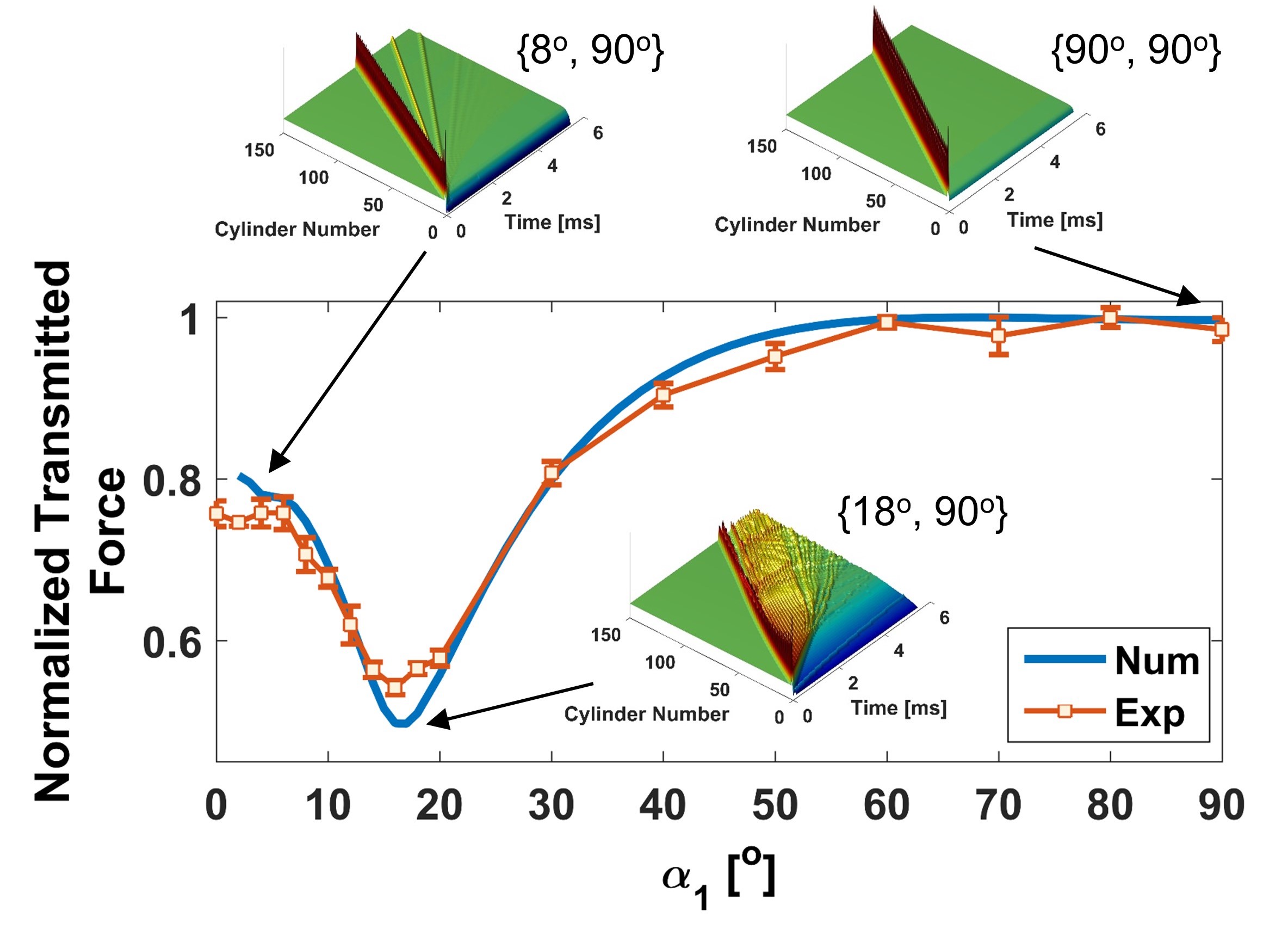}
\caption{\textbf{Force transmission through the dimer chain when given an input impulse.} Insets show the spatio-temporal maps of the cylinder velocity at angles, which support wave localization ($\alpha_1={90^{\circ},~8^{\circ}}$), and strong dispersion ($\alpha_1=18^{\circ}$).} 
\label{fig8}
\end{figure} 

In order to further validate these multiple wave localization and strong dispersion cases, we look at the force transmission through the chain for multiple dimer configurations. This is calculated by normalizing all the measured forces at the other end with respect to the force obtained in the case when $\alpha_1, \alpha_2=\{90^{\circ}, 90^{\circ}\}$, i.e., a homogeneous system. In Fig.~\ref{fig8}, we plot numerically and experimentally obtained normalized transmitted force. We clearly see complete force transmission in case of  $\alpha_1=90^{\circ}$, and corresponding numerically obtained spatio-temporal map of velocity confirming the solitary wave propagation (inset). However, as the angle is reduced, there is a steep decline in the force transmission. Interestingly, we see that there are certain contact angles (alternatively, stiffness coefficients ratios), for which, force transmission curve attains local minima. The first such point can be seen at about $\alpha_1=18^{\circ}$. The adjacent inset shows the spatio-temporal map of wave propagation. Clearly, dispersion effects are maximum here, leading to more than $40 \% $ loss in transmitted force in the chain with 40 cylinders. 

When we further reduce the angle, we reach a point again, which shows a local jump in force transmission. The corresponding inset indicates wave localization in the form of solitary wave for this case. This second localization occurs at about $\alpha_1=8^{\circ}$. This indeed makes sense because aforementioned numerical and experimental data showed that this $\kappa$ invokes wave localization in the system. As per analytical and numerical studies, any further reduction in the angle will lead to additional critical angles supporting wave localization, and reflecting as these local maxima in force transmission. This argument can be further extended by stating that there will also be additional angles showing local \textit{minima} in force transmission due to strong dispersion effects \citep{Jayaprakash2013}. 
However, it was not possible to validate any more of such points due to practical limitations on the test setup and contact sensitivity for small angles.    

\begin{figure}[t]
\centering
\includegraphics[scale=0.55]{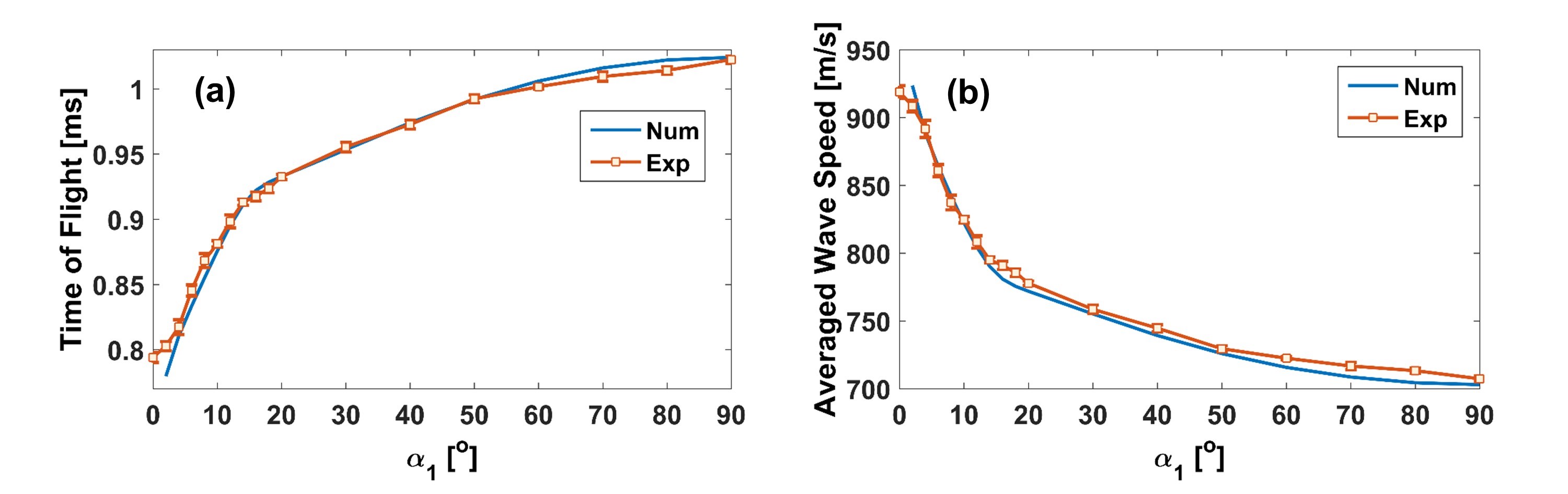}
\caption{(a). Time it takes for the wave front to reach the other end of the chain. (b) Average wave speed based on the time of flight.}
\label{fig9}
\end{figure} 

Figure~\ref{fig9}a shows the time taken by the input pulse to reach the other end of the dimer chain at several $\alpha_1$ values. In Fig.~\ref{fig9}b we show the average wave speeds obtained from corresponding time of flight data. We see an excellent match between numerical and experimental data, and these make sense as decreasing $\alpha_1$ results in higher system stiffness leading to faster wave propagation.

\section{Unique impact mitigation mechanism}
\begin{figure}[t]
\centering
\includegraphics[scale=0.55]{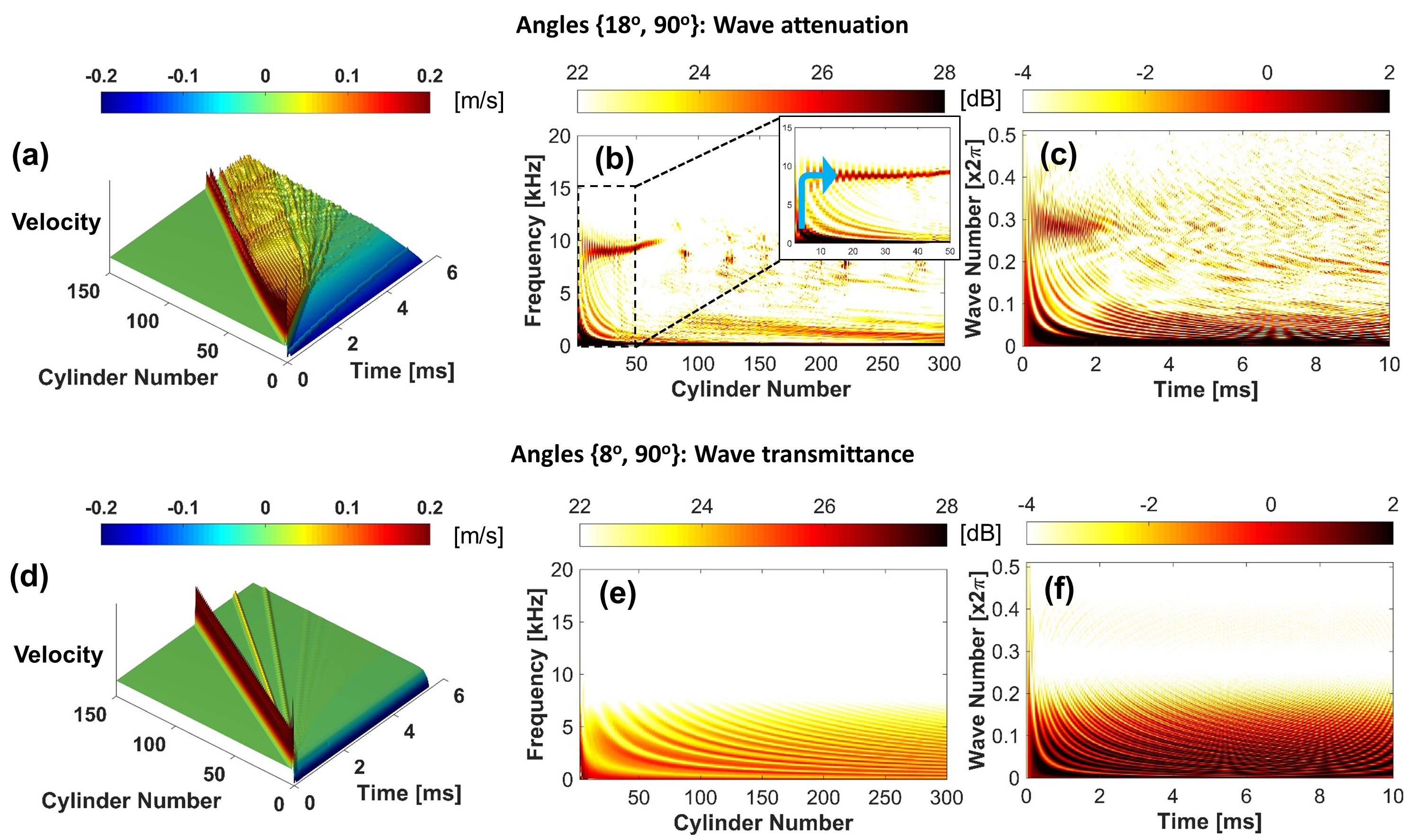}
\caption{\textbf{Energy redistribution in frequency and wavenumber domains.} (a) Spatio-temporal map of cylinder velocity at strong dispersion case. (b-c) Corresponding Fourier transformations in time and space domains. (d-f) The same at wave localization case.}
\label{fig10}
\end{figure} 

\begin{figure}[t]
\centering
\includegraphics[scale=0.55]{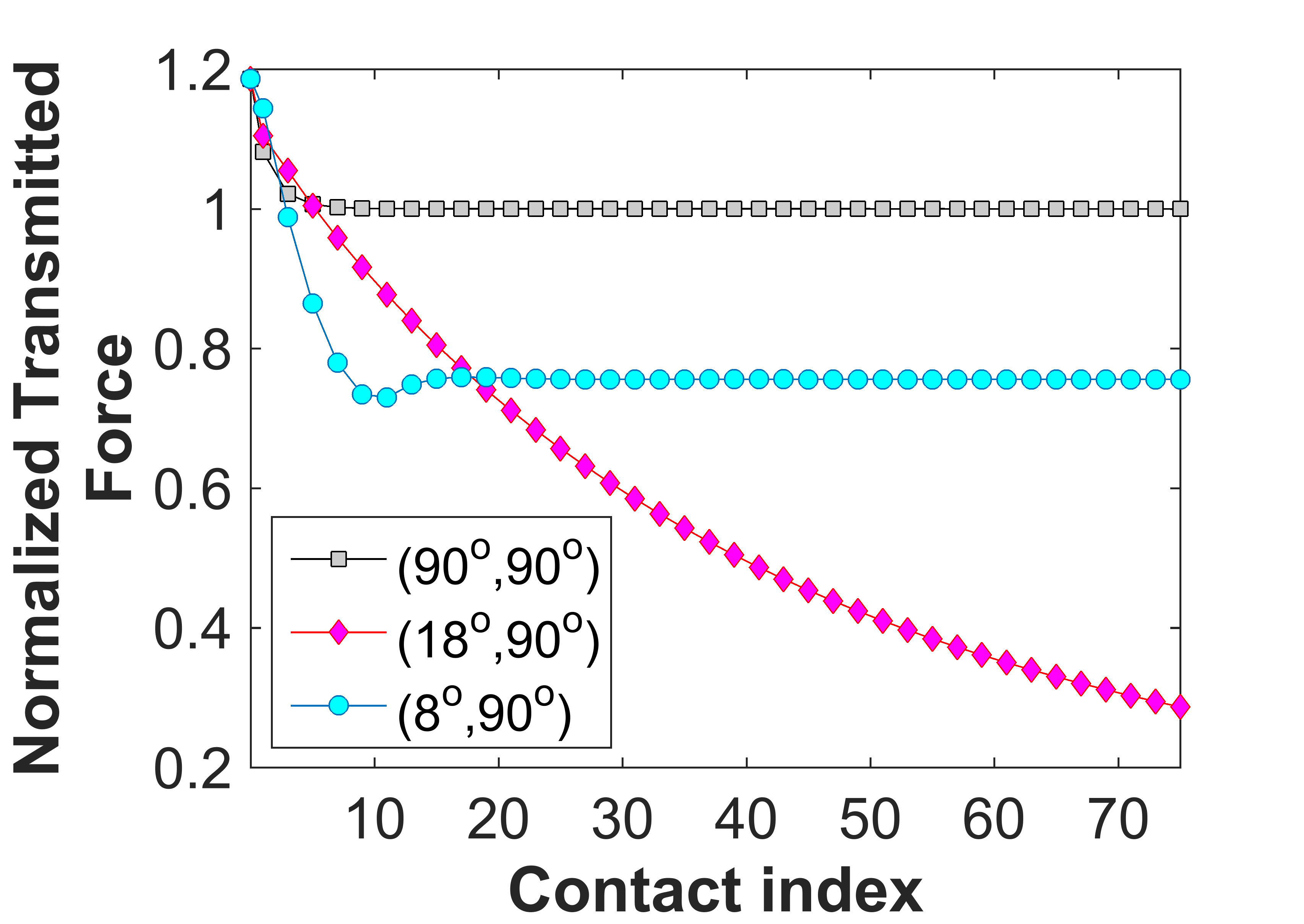}
\caption{Normalized transmitted force in the system as a function of dimer length (contact index)}
\label{fig11}
\end{figure} 

We have so far seen that the cylinder-based dimer system offers extreme flexibility in terms of achieving desired impulse transmission. This is due to the possibility of both localized as well as highly dispersive wave propagation in the same medium. Now we would like to dive deeper into the strong dispersion mechanism, which leads to drastic attenuation of impulse. To this end, we rely on the spectrum data of the wave signal. For the ease of distinction, we take two extreme cases, one with localization ($\alpha_1=8^{\circ}$), and the other with maximum dispersion ($\alpha_1=18^{\circ}$). In Fig.~\ref{fig10} we plot numerically obtained spatio-temporal maps for cylinder velocity, and its FFT in space and time domains. Figure~\ref{fig10}a depicts the spatio-temporal map of cylinder velocity as wave propagates in the medium for $\alpha_1=18^{\circ}$. Figure~\ref{fig10}b shows the frequency content in the traveling wave along the space. We see an interesting redistribution of energy at the starting of the chain, in which, the energy transfers from low frequency to high frequency domain (inset in Fig.~\ref{fig10}b). This passive ability of relocating impact energy is very distinctive feature of this dimer (periodic) system. In Fig.~\ref{fig10}c, we plot wavenumber against time. In this strongly nonlinear system, we see energy redistribution even in the wavenumber domain. Clearly wave disintegrates to small length scales (or high wavenumber) as time proceeds. This hints a turbulence-like cascading mechanism in the dimer system.  However, in case of localization at $\alpha_1=8^{\circ}$ as seen from  Fig.~\ref{fig10}d-f, there is no such energy redistribution mechanism at work. 

Finally, for more practical side of the study, we show how long the system should be in order to achieve desired attenuation in force. Figure~\ref{fig11} depicts the normalized transmitted force (with respect to transmitted force in case of homogeneous chain) for three dimer configurations. One can see that the cases where wave localization is involved, i.e., $\{8^{\circ},90^{\circ}\}$ and $\{90^{\circ},90^{\circ}\}$, transmitted force quickly settles to a saturated value. In these cases, there is a slight reduction in the force before localized traveling wave is fully established. However, due to the unique impact mitigation mechanism in case of  $\{18^{\circ},90^{\circ}\}$, one sees strong dispersion as the wave travels along the chain, and thus it leads to drastic attenuation of impact force. It would be interesting to further examine these wave transmission/attenuation characteristics in presence of dissipation. Initial studies on sphere-based granular systems show that qualitative nature of these mechanisms may depend on the strength of dissipation in the system \citep{Wang2015}.

\section{Conclusion and future directions}
We have presented a novel phononic crystal based on cylinder particles in order to manipulate the impulse wave in the system. A periodic arrangement of such device, i.e., a `dimer', is shown to be supporting two extremes of elastic wave propagation in the medium. Due to the rotational tunability of such a system, one can \textit{in-situ} change the contact angles between the cylinders, and thus an impulse on the system can either be localized and transmitted without any attenuation, or it can be highly dispersed leading to strong attenuation. We have employed analytical, numerical, and experimental techniques to validate these wave dynamics, which stem from the interplay of dispersion 
and nonlinear effects in the system.  

We have shown that wave localization occurs for multiple (a countable infinity) dimer configurations. Each such configuration is characterized by the stiffness coefficient ratio $\kappa$. The localized, solitary wave, was found to have symmetry in the time-history of contact forces, instead of that of velocity, deviating from the conventional notion of solitary wave in Hertz type granular system.

There are also multiple dimer configurations that maximize dispersion effects on the wave instead of localization. These result in strong attenuation of impact force. Fundamentally, these cases show unique LF-HF (low-to-high frequency) energy transfer mechanism. It would be interesting to explore how the presence of losses (dissipation) in the system can complement/destroy this dissipation-independent energy transfer mechanism. For example, the effects such as friction, plasticity, visco-elasticity, and viscous drag could be considered for the same. Hence, this opens up new possibilities for optimizing the structure for impact mitigation purposes by tailoring several material parameters. Moreover, turbulence-like cascading across different length scales in an ordered system can be of interest to researchers.
Finally, this study can motivate the design of a general class of discrete systems using various nonlinear interactions for tailoring impulse wave propagation. 
  
\section*{Acknowledgments}  
We gratefully thank Prof. P. Kevrekidis, University of Massachusetts Amherst, and A. Vakakis, University of Illinois Urbana-Champaign for insightful discussions. J.Y. thank the support of the U.S. ONR (N000141410388) and Korea's ADD. E.K. acknowledges the support of research funds for newly appointed professors at the Chonbuk National University in 2016 and the support from Leading Foreign Research Institute Recruitment Program through the National Research Foundation of Korea funded by the Ministry of Science, ICT and Future Planning (2011-0030065).
\section*{References}  
\bibliography{mybib}

\end{document}